\documentclass[prl,amsmath,amsfonts,amssymb,twocolumn,superscriptaddress,showpacs]{revtex4}

\usepackage{graphicx,psfrag}
\usepackage{dcolumn}
\usepackage{bm}
\usepackage{mathbbol}
\usepackage{makeidx}               
\usepackage{graphicx}              
\usepackage[bookmarksnumbered,colorlinks,plainpages,citecolor=blue,urlcolor=blue]{hyperref}
\usepackage[rflt]{floatflt}
\usepackage{subfigure}
\usepackage{caption2}
\usepackage{latexsym}
\usepackage{amsmath}
\usepackage{amsfonts}
\usepackage{amssymb}
\usepackage{enumerate}

\newcommand{\uck}[1]{\o}

\newcommand{\ket}[1]{\mbox{$|#1\protect\rangle$}}

\newcommand{\bra}[1]{\mbox{$\protect\langle#1|$}}

\def\beq{\begin{equation}}
\def\eeq{\end{equation}}
\def\bea{\begin{eqnarray}}
\def\eea{\end{eqnarray}}

\begin{document}

\begin{titlepage}
\date{\today}
\title{Quantum Gaussian Channels with Weak Measurements}

\author{Boaz Tamir}
\email{boaz_tamir@post.bezalel.ac.il}
\affiliation{\mbox{Faculty of Interdisciplinary Studies, Bar-Ilan University, Ramat-Gan, Israel}}

\author{Eliahu Cohen}
\email{eliahuco@post.tau.ac.il}
\affiliation{\mbox{School of Physics and Astronomy, Tel Aviv University, Tel Aviv, Israel}}


\maketitle

\section{Abstract}

In this paper we perform a novel analysis of quantum Gaussian channels in the context of weak measurements. Suppose Alice sends classical information to Bob using a quantum channel. Suppose Bob is allowed to use only weak measurements, what would be the channel capacity? We formulate weak measurement theory in these terms and discuss the above question.

\section{I. Introduction and Motivation}


Gaussian channels were first presented in Shannon's work \cite{Shannon}. They appear naturally in classical communication theory \cite{Cover}. A Gaussian channel is a channel where the output $Y$ is the sum of the input $X$ and a noise $Z$, where the noise is drawn randomly from a Gaussian distribution usually with $0$ mean and $\sigma^2$ variance \cite{Cover}. In this work the noise is introduced through the weak measurement process at the end point of the channel, rather than as a result of some external process.\\


The capacity of quantum Gaussian channels was calculated in \cite{HolevoG1,HolevoG2} and was further studied in \cite{G3,G4,G5}. In this paper, we will rely on these works by choosing particular cases of the above theory, but at the same time, we will adopt rather different perspective and applications from the known ones.

Weak measurements were first presented by Aharonov {\it et al.} in \cite{AAV}. In a nutshell, a measurement is weak if the measuring apparatus, which contains a quantum element we call a `needle', is weakly coupled to the quantum system we wish to measure, and therefore yields a set of eigenvalues covered with random noise. The needle is usually normally distributed with high variation.

So far, weak measurements were not discussed in this language of quantum channel capacity. We believe that such a discussion can clearly draw the line between strong and weak measurements in quantum mechanics. By `weak measurement' we refer to a lower bound on the variance of the needle (see below), therefore all the above is true for a large set of measurements strength.



Our main result can be stated as follows: let $\{x_i\}$ be the set of letters used by Alice. Let $X$ be a random variable describing the distribution of letters Alice tries to send. Suppose Alice is employing a quantum channel, sending a density matrix $\rho_i$ for each of the letters $x_i$. Next, Bob is performing a weak measurement process \cite{Aharonov} of some Hermitian operator $\hat{A}$ at the end point of the channel. We assume therefore that the $\{x_i\}$ are the eigenvalues of $\hat{A}$, this should be part of the protocol Alice and Bob share. Let $Z$ denote the normal Gaussian distribution $N(0, \sigma^2)$ of Bob's measurement needle before the weak coupling. Then the maximal rate of transmission through the channel will be:

\begin{eqnarray}
\max_{X(\hat{A})} H(X:X+Z),
\end{eqnarray}

\noindent where $H(X:X+Z)$ is the classical mutual information between the two classical random variables, and the maximum is taken over all discrete probability distributions $X$ on the set $\{x_i\}$ of eigenvalues of $\hat{A}$. The above formula resembles the one presented by Shannon for a Gaussian channel \cite{Shannon}, however here it has a different meaning.








\noindent Moreover, let
\begin{eqnarray}
\max_{X(\hat{A})}E(X^2)= P_{\hat{A}}.
\end{eqnarray}

\noindent Then the capacity of the weak channel will be bounded from above by:
\begin{eqnarray}
\label{main}
\max_{X(\hat{A})} H(X:X+Z) < \frac{1}{2}log \left(\frac{P_{\hat{A}}} {\sigma^2}+1\right).
\end{eqnarray}

\noindent This is the main result of the paper which, to our understanding, is both novel and insightful. Note that in the classical case all signals passing the channel are {\it a priori} bounded by the power of the signal. Here, the maximal variance of all probability distributions on $\hat{A}$ takes the place of the power. This calls for a more general definition of a weak channel where one can use any Hermitian matrix in the weak measurement process. Note also that the upper bound in $\ref{main}$ is achieved in the classical case. The difference between the classical and quantum bound is due to the definition of the channel. In classical channels the Gaussian noise is presented during the passage of information from the transmitter to the receiver. In the Quantum weak Gaussian channel (QWGC) case the `noise' is presented at the last step of the communication, during the process of weak measurement. In particular, the codewords in the classical case are produced by sampling from a normal distribution, whereas in the quantum case the codewords are sampled from a finite discrete ensemble of density matrices (see also the proof of part (a) of the main theorem in section III) hence the difference in the bound. The method used in this paper combines arguments from discrete channel theory with arguments from continuous Gaussian channel theory \cite{Cover}.

Two factors therefore define the upper bound on the rate of information flow of QWGC; the maximal variance of $\hat{A}$, and $\sigma^2$, the variance of the needle.


The rest of the paper is organized as follows: in section II we introduce some necessary weak measurements preliminaries and prove two useful lemmas. In section III we present the notion of quantum weak gaussian channels and prove the main result of the paper.

\section{II. Weak Measurement Preliminaries}

Weak measurement should be treated as a generalization of quantum
strong measurement. In weak measurement theory both the system and
the measuring needle are quantum systems \cite{Aharonov,Tamir,ACE}. Weak
measurement consists of two steps. In the first step we weakly couple
the quantum measurement device to the quantum system via a von Neumann interaction Hamiltonian. In the second step we `strongly' measure the needle. The collapsed state
of the measurement device is referred to as the outcome of the weak
measurement process. For a measurement to be weak, the standard deviation
of the measurement needle should be larger than the difference between
the eigenvalues of $\hat{A}$. We will now describe this process in
details. The procedure resembles the von Neumann scheme for performing `strong' measurements \cite{Neumann}, however, here we use a very weak entanglement between
the system and the measurement device (see also \cite{Aharonov,Tamir,ACE}).

Let $|\phi_{d}\rangle$ denote the wave function of the measurement
device. When represented in the position basis it will be written
as:

\begin{equation}
|\phi\rangle=|\phi_{d}\rangle=\int_{y}\phi(y)|y\rangle dy,
\end{equation}

\noindent where $y$ is the position variable of the measuring needle.
Let $\hat{Y_{d}}$ be the position operator such that $\hat{Y_{d}}|y\rangle=y|y\rangle$
(here, we use $\hat{Y}_d$ to distinguish the operator $\hat{Y}_d$ from
its eigenvector $|y\rangle$ and eigenvalue $y$, the subscript $d$ is used for measuring device). We will also assume that initially $\phi(y)$ behaves normally around $0$ with some variance $\sigma^{2}$:

\begin{equation}
\phi(y)=(2\pi\sigma^{2})^{-\frac{1}{4}}e^{-y^{2}/4\sigma^{2}}.
\end{equation}

\noindent We will later (strongly) measure $|\phi_{d}\rangle$, {\it i.e.}
collapse the device's needle to get a value which is the weak measurement's
outcome.

Let $S$ denote our system to be measured. Suppose $\hat{A}$ is an
Hermitian operator on the system $S$. Suppose $\hat{A}$ has $N$
eigenvectors $|x_{i}\rangle$ such that $\hat{A}|x_{i}\rangle=x_{i}|x_{i}\rangle$.

Consider the general state vector $|\psi\rangle$ expressed in the
eigenbasis of $\hat{A}$:

\begin{equation}
\label{ro}
|\psi\rangle=\sum_{i}\alpha_{i}|x_{i}\rangle.
\end{equation}

\noindent Consider the interaction Hamiltonian $\hat{H}_{\textrm{int}}$
(\cite{Tamir,ACE}):

\begin{equation}
\hat{H}=\hat{H}_{\textrm{int}}=g(t)\hat{A}\otimes\hat{P_{d}}.
\end{equation}

\noindent Here $g(t)$ is a coupling impulse function satisfying:

\begin{equation}
\int_{0}^{T}g(t)dt=1, \label{eq:cif}
\end{equation}

\noindent where $T$ is the coupling time and $\hat{P_{d}}$ is the
operator conjugate to $\hat{Y_{d}}$ such that $[\hat{Y_{d}},\hat{P_{d}}]=\imath\hbar$. For the measurement to be considered weak, the coupling strength should be much smaller than the standard error $\sigma$ of the measurement needle.

We shall start the measurement process with the vector:

\begin{equation}
|\psi\rangle\otimes|\phi(y)\rangle,
\end{equation}

\noindent in the product space of the two systems. Then we apply the
following time evolution based on the weak measurement Hamiltonian \cite{Tamir,ACE}:

\begin{equation}
e^{-\imath\hat{H}t/\hbar}|\psi\rangle\otimes|\phi(y)\rangle.
\end{equation}

\noindent It is easy to see that on each of the vectors $|x_{i}\rangle\otimes|\phi(y)\rangle$
the Hamiltonian $\hat{H}$ takes $\hat{Y}_{d}$ to $\hat{Y}_{d}+x_{i}$,
(Heisenberg evolution):

\begin{align}
\hat{Y_{d}}(T)-\hat{Y_{d}}(0) & =\int_{0}^{T}dt\frac{\partial\hat{Y_{d}}}{\partial t}\nonumber \\
 & =\int_{0}^{T}\frac{\imath}{\hbar}[\hat{H},\hat{Y_{d}}]dt=x_{i}
\end{align}

\noindent (see \cite[section 8.4]{Peres}). The corresponding transformation
of the coordinates of the wave function is:

\begin{equation}
e^{-\imath\hat{H}T/\hbar}|\psi\rangle\otimes|\phi\rangle=\sum_{i}\alpha_{i}|x_{i}\rangle\otimes|\phi(y-x_{i})\rangle. \label{eq:weakmeas}
\end{equation}

\noindent The above wave functions $|\phi(y-x_{i})\rangle$ have high variance and
therefore they overlap each other. The higher the variance, the weaker
the measurement process. If these normal wave functions do not overlap
then the measurement is strong. Therefore we can control the measurement process
by the choice of the variance.




Let $\hat{U} = e^{-\imath\hat{H}T/\hbar}$, let $\rho$ be a density matrix on the main system $Q$, let $\ket{\phi} \bra{\phi}$ be the density matrix for the needle space of Bob, let $\hat{P}_y= \ket{y} \bra{y}$. We can then write:

\[ p(Y=y)dy = tr(\hat{P}_y  U (\rho \otimes (\ket{\phi} \bra{\phi})) U^\dagger \hat{P}_y^\dagger)dy \]

\noindent The following two simple lemmas are immediate and are needed for the definition of $H_{\hat{A}}^{W,\rho}(Y)$.

\textbf{Lemma II.1}

If $\rho$ is pure, $\rho = \ket{\psi} \bra{\psi}$ (where $\ket{\psi}$ as in Eq. $\ref{ro}$ above), then

\[ p(y_1\leq Y \leq y_2) = \sum_i |\alpha_i|^2 \frac{1}{\sqrt{2\pi} \sigma} \int _{y_1}^{y_2} e^{-(z-x_i)^2/2\sigma^2}dz .\]

\textbf{Proof:} See \cite{Vaidman} for example. $\blacksquare$

The distribution of $Y$ is multi-normal with coefficients $|\alpha_i|^2$ and means $x_i$.

\textbf{Lemma II.2}

Under the above assumptions; if $\rho$ ia a density matrix, then




\begin{equation}
\label{entropy}
p(y_1\leq Y \leq y_2) = \sum_i tr(\rho \hat{P}_{\ket{x_i}})\frac{1}{\sqrt{2\pi} \sigma} \int _{y_1}^{y_2} e^{-(z-x_i)^2/2\sigma^2}dz
\end{equation}

\noindent where $\hat{P}_{\ket{x_i}}$ is the projective operator $\ket{x_i}\bra{x_i}$ and $Z_i \sim N(x_i,\sigma^2)$, is a normal distribution with mean $x_i$ and variance $\sigma^2$.

\textbf{Proof:} Use a simple Bayesian argument and the fact that $tr(\rho \hat{P}_{\ket{x_i}})$ is the probability that the $\hat{P}_{\ket{x_i}}$ projective measurement yields $x_i$. $\blacksquare$



\noindent \textbf{Notation:} Denote the entropy of the above multi-normal distribution (Eq. $\ref{entropy}$) on $Y$ by $H_{\hat{A}}^{W, \rho}(Y)$.

\section{III. Quantum Weak Gaussian Channels}

In this section we present and analyze QWGC and prove the main theorem of the paper. QWGCs are quantum channels where in the last step we use weak measurements instead of `strong' POVM measurements.

Suppose Alice is sending classical information to Bob using a quantum channel \cite{Nielsen}. Let $X$ denote a random variable such that $p(X=i) = p_i$. With probability $p_i$ Alice will pick a density matrix $\rho_i$ out of a given ensemble of density matrices. She will then send $\rho_i$ to Bob using a quantum (physical) channel. Define the density matrix:

\[ \rho =\sum_i p_i\rho_i .\]

At the output of the channel Bob is using a weak measurement device trying to guess the value of the random variable $X$. Bob is using an Hermitian operator that has the same eigenvalues as Alice's letters. The measurement's results yield a random variable $Y$.
To compute the mutual information between Alice and Bob we could use a Holevo type of bound. However, since Bob is using a particular measurement (strong projective on the continuous variable $Y$) we can compute $H_{\hat{A}}^{W, \rho}(X:Y)$ explicitly using the definitions of section II above:

\[ H_{\hat{A}}^{W, \rho}(X:Y) = H_{\hat{A}}^{W, \rho}(Y) - \sum_i p_iH_{\hat{A}}^{W, \rho_i}(Y)\]

In the following lemma we compute the maximal mutual information between Alice and Bob for all distributions $\{p_i\}_i$ and all densities $\{\rho_i\}_i$. In the main theorem that follows we will show that this maximal mutual information is also the capacity of the channel.

\textbf{Lemma III.1}

\[ \max_{p_i,\rho_i} H_{\hat{A}}^{W, \rho}(X:Y) = \max_{X(\hat{A})} H(X:X+Z) \]

\noindent where the R.H.S of the above equation is the maximal mutual information of two classical random variables $X(\hat{A})$ and $X(\hat{A})+Z$ where $X(\hat{A})$ is any discrete random variable on $\{x_i \}$, the eigenvalues of $\hat{A}$, and $Z$ is the normal random variable centered around $0$ with variance $\sigma^2$.

\textbf{Proof:}

\[ \textbf{a)} \hspace{5mm} \max_{p_i,\rho_i} H_{\hat{A}}^{W, \rho}(X:Y) \leq \max_{X(\hat{A})} H(X:X+Z). \]

\noindent Suppose we are given $\{p_i\}$ and $\{\rho_i\}$. For each density matrix $\rho_i$, the distribution of $Y$ is multi-normal (see lemma II.2 above) and therefore, by the concavity of entropy (see 11.3.5 of \cite{Cover}):

\begin{equation}
H(Z)= H(N(0,\sigma^2)) \leq  H_{\hat{A}}^{W, \rho_i}(Y),
\end{equation}

\noindent for all $i$. Define the following random variable $X$ on the set $\{x_i\}_1^d$ by $p(X=x_i)=tr(\rho M_{\ket{x_i}})$, then

\begin{equation}
H_{\hat{A}}^{W, \rho}(Y) - \sum p_i H_{\hat{A}}^{W, \rho_i}(Y) \leq H(X+Z)-H(Z),
\end{equation}

\noindent where we have used the fact that $H(N(0,\sigma^2))=H(N(x_i,\sigma^2))$ for all $i$. Conclude now by recalling that $H(X:X+Z)=H(X+Z)-H(Z)$ \cite{Cover}.

\[ \textbf{b)} \hspace{5mm} \max_{X(\hat{A})} H(X:X+Z) \leq \max_{p_i,\rho_i} H_{\hat{A}}^{W, \rho}(X:Y).  \]

For each distribution $X(\hat{A})$ on the L.H.S we find a density $\rho$ on the R.H.S. Given a random variable $X$, such that $p(X=x_i)=p_i$, we define:

\[ \rho_i = \ket{x_i} \bra{x_i}. \]

\noindent Then $H_{\hat{A}}^{W, \rho_i}(Y)= H(N(x_i,\sigma^2))$. Let  $\rho = \sum p_i \rho_i$, then:

\[ H(X:X+Z)= H_{\hat{A}}^{W, \rho}(X:Y), \]

\noindent hence the above inequality. $\blacksquare$

\noindent \textbf{Main Theorem:}


The rate of transmission of classical information in the QWGC, that is, the channel between Alice and Bob is:


\begin{equation} \label{Rate}
R =  \max_{p_i,\rho_i} \{ H_{\hat{A}}^{W, \rho}(Y) - \sum p_i H_{\hat{A}}^{W, \rho_i}(Y) \}.
\end{equation}

\textbf{Proof:} Denote:

\[ \chi^W_{\hat{A}} = \max_{p_i,\rho_i} \{ H_{\hat{A}}^{W, \rho}(Y) - \sum p_i H_{\hat{A}}^{W, \rho_i}(Y)\}. \]

\noindent We will prove that:

\[ (a) \hspace{5mm} R \geq \chi^W_{\hat{A}} \]

\[ (b) \hspace{5mm} R \leq \chi^W_{\hat{A}} \]

To prove (a) we use a reduction to classical argument. To prove (b) we use the subadditivity of entropy.

\textbf{Proof of} (a): Given an {\it a priori} distribution $p$, Alice can send one of a $2^{nR}$ predetermined $n$-products:

\[ M_1 = \rho_1^1 \otimes...\otimes \rho_n^1 \]

\[ M_{2^{nR}} = \rho_1^{2^{nR}} \otimes...\otimes \rho_n^{2^{nR}}, \]

\noindent where $M_i$ stands for the $i$-th codeword. Note that $\rho_i^j$ can be any density matrix over $\{\ket{x_i}\}_{i=1}^d$, hence we can choose:

\[\rho_i^j= \ket{x_i^j}\bra{x_i^j},\]

\noindent where each of the $M_i$ codewords is $\epsilon$-typical according to the {\it a priori} distribution defined by $X$. Bob will weakly measure the product by coupling $n$ normal needles and strongly measuring all of them. By a standard classical argument we can separate $H(X:X+Z)$ such tensors \cite{Cover}. Thus:

\[ R \geq H(X:X+Z) \]

\noindent Maximizing now over all {\it a priori} distributions $p$ on $X$, we get:

\[ R \geq \max_{X(\hat{A})} H(X:X+Z). \]

\noindent By lemma III.1 above:

\begin{equation}
R \geq \max_{p_i,\rho_i} \{H_{\hat{A}}^{W, \rho}(Y) - \sum p_i H_{\hat{A}}^{W, \rho_i}(Y) \},
\end{equation}
















\noindent which concludes the proof of (a).

\textbf{Proof of} (b): We follow the argument in \cite{Nielsen} with some necessary adjustments. Suppose Alice and Bob are using one of $2^{nR}$ ($n$-product) codewords:

\[ M_1 = \rho_1^1 \otimes ... \otimes \rho_n^1 = \overline{\rho}^1, \]

\[ M_{2^{nR}} = \rho_1^{2^{nR}} \otimes ... \otimes \rho_n^{2^{nR}}= \overline{\rho}^{2^{nR}} \]

\noindent (this time the densities $\rho_i^j$ are not necessarily pure).

\noindent Let $M$ be the classical random variable defined by using each of the $2^{nR}$ codewords with equal probability. Suppose Bob weakly couples a set of $n$ needles on  $\overline {Y}= Y^{\otimes n}$, where on each subspace he uses the procedure discussed in the previous section to a perform weak measurement with a single needle.

In what follows, we will first bound the global mutual information $H(M:\overline {Y})$:

\[ H(M:\overline {Y}) \leq n \cdot \chi^W_{\hat{A}},\]

\noindent and then we will conclude by relating it to Bob's error probability using the Fano inequality \cite{Cover}.



We begin with some definitions. In the previous section we defined the entropy $H_{\hat{A}}^{W,\rho} (Y)$ (Eq. $\ref{entropy}$) for a general density matrix $\rho$ on $X = X(\hat{A})$. We now extend the definition to the product spaces $\overline {X}= (X(\hat{A}))^{\otimes n}$. Consider the above codewords density matrix $\overline{\rho}$ on $\overline {X}$:
\begin{equation}
\overline{\rho} = \sum_{i=1}^{2^{nR}} \frac{1}{2^{nR}} \overline{\rho}^i,
\end{equation}

\noindent where $\overline{\rho}^i$ is the codeword $M_i$ as defined above. Following the weak coupling of measurement needles we can define $H_{\hat{A}}^{W,\overline{\rho}} (\overline{Y})$ to be the entropy of the following distribution on $\overline{Y}$:

\[ p\{ (y_{1,0},...,y_{n,0}) \leq (Y_1,...,Y_n) \leq (y_{1,1},...,y_{n,1})\}\]
\[=\sum _{x_{i_1},...,x_{i_n}} tr(\overline{\rho} M_{\ket{x_{i_1}}} \cdot \cdot \cdot M_{\ket{x_{i_n}}}) \cdot \]

\[\cdot \int_{y_{1_0}}^{y_{1_1}} \frac{1}{\sqrt{2\pi} \sigma} e^{-\frac{(y_1-x_{i_1})^2} {2 \sigma^2}} dy_1 \cdot \cdot \cdot \int_{y_{n_0}}^{y_{n_1}} \frac{1}{\sqrt{2\pi} \sigma} e^{-\frac{(y_n-x_{i_n})^2} {2 \sigma^2}} dy_n,\]

\noindent where $i_1,...,i_n$ are in $\{1,...,d\}$. Now since $M$ and $\overline{Y}$ behave as classical random variables we can write:











\begin{equation}
\label{a}
H(M:\overline{Y})= H_{\hat{A}}^{W,\overline{\rho}} (\overline{Y}) -\sum_{i=1}^{2^{nR}}\frac{1}{2^{nR}} H_{\hat{A}}^{W,\overline{\rho}^i} (\overline{Y}).
\end{equation}


To continue the proof of the theorem we shall need the following Lemma.

\textbf{Lemma III.2}
\begin{equation}
H(M:\overline{Y}) \leq n \cdot \chi^W_{\hat{A}}.
\end{equation}





\textbf{Proof:} Consider the following density matrix on $X^j(A)$:
\begin{equation}
\label{jcoefficient}
{\rho_j} = \sum_{i=1}^{2^{nR}} \frac{1}{2^{nR}} {\rho}_j^i.
\end{equation}
\noindent Note that $\rho_j$ is a column in the matrix $(\rho_j^i)$, and can be computed by tracing out the rest of the coordinates of $\overline{\rho}$. By the subadditivity on $\overline{Y}$ (see \cite{Nielsen} chapter 11.3.4):

\begin{equation}
\label{b}
H_{\hat{A}}^{W,\overline{\rho}} (\overline{Y})\leq \sum_{j=1}^n H_{\hat{A}}^{W,{\rho}_j} ({Y_j})
\end{equation}

\noindent Subadditivity also implies that on the tensor product $\overline {\rho}^i=\rho_1^i \otimes...\otimes \rho_n^i$ (a row in the matrix $\rho_j^i$) we can write:

\begin{equation}
\label{c}
H_{\hat{A}}^{W,\overline{\rho}^i} (\overline{Y})= \sum_{j=1}^n H_{\hat{A}}^{W,{\rho}_j^i} ({Y}_j).
\end{equation}

Combining Eqs. $\ref{a}$, $\ref{b}$ and $\ref{c}$ we can write:

\begin{equation}
\label{d}
H(M:\overline{Y}) \leq \sum_{j=1}^n \{ H_{\hat{A}}^{W,{\rho}_j} ({Y}_j)     - \sum_{i=1}^{2^{nR}}\frac{1}{2^{nR}} H_{\hat{A}}^{W,{\rho}_j^i} ({Y}_j)\}
\end{equation}

\[ \leq n \max_{\rho = \sum p_k \rho_k} \{ H_{\hat{A}}^{W, \rho}(Y) - \sum p_k H_{\hat{A}}^{W, \rho_k}(Y)\}. \]

\noindent The last inequality is due to the fact that $\rho_j$ is a sum of density matrices on one coordinate (the $j$-th, see Eq. $\ref{jcoefficient}$). $\blacksquare$

We can now complete the proof of (b). Since Alice is sending each $M_i$ with equal probability $\frac{1}{2^{nR}}$, then the probability $P_{err}$ that Bob is mistaken is the average:

\[ P_{err} = \sum_i  \frac{1}{2^{nR}} P_{err}^i, \]

\noindent where $P_{err}^i$ is the probability for Bob (using weak measurements) to make a mistake, given that Alice had sent $M_i$ (we do not have to know how to compute $P_{err}^i$). Since $R$ is the classical rate of transmission (Eq. \ref{Rate}), it is assumed that $P_{err}$ can be arbitrary small for large enough $n$. By the Fano inequality on the random variable $M$ \cite{Cover}, we know that:

\[ H(P_{err}) + P_{err}\cdot log(d^n) \geq H(M/\overline{Y}), \]

\noindent where we have used the fact that measuring $\overline{Y}$ yields no more than $d^n +1$ values for $M$. Hence,

\[ n P_{err} \cdot log(d) \geq H(M)-H(M:\overline{Y})-H(P_{err})\]
\[= nR-H(M:\overline{Y})- H(P_{err}). \]

\noindent Since $H(M:\overline{Y}) \leq n \chi^W_{\hat{A}}$ (by Eq. $\ref{d}$ above) we have (for large enough $n$):

\[ P_{err}\geq \frac{R-\chi^W_{\hat{A}}}{log(d)}, \]

\noindent thus concluding that $R\leq \chi^W_{\hat{A}}$, otherwise $P_{err}>0 $.  $\blacksquare$

We will now show a simple bound on the rate of transmission using the above theorem. Let the random variables $X$ on the space of eigenvalues of $\hat{A}$, $Z$ and $Y$ on the needles space, be defined as above ($Z$ before the weak measurement and $Y$ following the weak measurement). Let $E(X)$ denote the expectation value of $X$. Since $E(Z)=0$ we can write:

\[E(X+Z)^2 = E(X^2) + E(Z^2). \]

\noindent  It is well known that the normal distribution maximize the entropy of all distributions of the same variance (see \cite{Cover} chapter 9), therefore:

\[ H(X+Z)< \frac{1}{2}log \left( 2\pi e(E(X^2) +\sigma^2)\right),\]

\noindent where $E(Z^2)=\sigma^2$  and the log function has base 2. Also

\[ H(Z) = \frac{1}{2} log\left( 2 \pi e \sigma^2\right).\]

\noindent Therefore
\begin{eqnarray}
H(X:X+Z)< \frac{1}{2}log \left(\frac{E(X^2) +\sigma^2}{\sigma^2}\right).
\end{eqnarray}

\noindent Moreover, let
\begin{eqnarray}
\max_{X(\hat{A})}E(X^2)= P_{\hat{A}}.
\end{eqnarray}

\noindent Then the capacity of the weak channel will be bounded from above by:
\begin{eqnarray}
\max_{X(\hat{A})} H(X:X+Z) < \frac{1}{2}log \left(\frac{P_{\hat{A}}} {\sigma^2}+1 \right).
\end{eqnarray}


\noindent \section{IV. Discussion}

We have defined and analyzed the channel capacity of the QWGC. Very naturally, it turns out that the maximal capacity depends on the maximal second moment of the operator used for creating Alice's letters divided by the variance of Bob's measuring needle.

One application of the theorem could be as follows: suppose Eve is eavesdropping a channel between Alice and Bob. To reduce the amount of interference she might want to use weak measurements. Suppose Eve knows the protocol, that is, the set of letters Alice and Bob are using. Eve is introducing a noise into the channel and Bob is trying to measure the output with a POVM measurement.

The noise introduced by Eve can be described by a trace preserving operator $\mathcal{E}_{\hat{A}}^W(\rho)$ on $\rho$ as discussed below (see Fig. 1);




\setlength{\unitlength}{0.75mm}
\begin{picture}(70,70)(-30,-20)
\put(0,0){\framebox(30,30)}
\put(5,23){\makebox(0,0)[bl]{$Q$}}
\put(5,3){\makebox(0,0)[bl]{$Y$}}
\put(-25,23){\makebox(0,0)[bl]{$\rho$}}
\put(-32,3){\makebox(0,0)[bl]{$\ket{\phi}\bra{\phi}$}}
\put(38,28){\makebox(0,0)[bl]{$\mathcal{E}_A^W(\rho)=\int p(y)\rho_y dy$}}
\put(40,7){\makebox(0,0)[bl]{$\hat{P_y}$}}
\put(-20,5){\line(1,0){20}}
\put(-20,25){\line(1,0){20}}
\put(30,25){\line(1,0){30}}
\put(30,5){\line(1,0){5}}
\put(30,5){\line(1,0){30}}
\put(45,5){\line(1,0){10}}
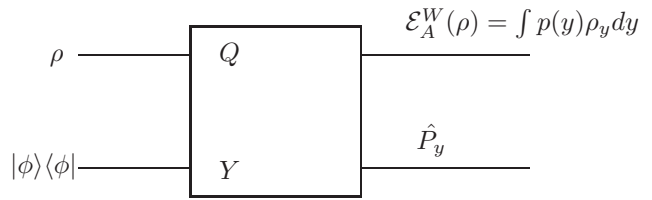
\put(-25,-15){\makebox(0,0)[bl]{{\bf Fig.1.}  Schematic illustration of the `weak channel'.  }}
\end{picture}\\


Suppose the system $Q$ is described by the density matrix $\rho$, and the system $Y$ (the needle) is described by the pure (and continuous) Gaussian vector state $\ket{\phi}$:

\[ \ket{\phi(y)} = (2\pi\sigma^{2})^{-\frac{1}{4}}\int e^{-y^{2}/4\sigma^{2}} \ket{y}dy. \]




First, Eve weakly couples $\rho$ and $\ket{\phi}\bra{\phi}$ by using a unitary operator $ U= U(\hat{A})$:

\[ U (\rho \otimes \ket{\phi}\bra{\phi}) U^\dagger \]


Next Eve acts on the product state by a set of projective operators $\hat{P_y}=\ket{y} \bra{y}$, this corresponds to a strong measurement of the needle's space which slightly alters the state of the measured system but with no collapse (see details in section II).






\noindent Using the standard operator sum representation \cite{Nielsen} we can define:

\begin{eqnarray}
\mathcal{E}_{\hat{A}}^W(\rho) = \int \hat{P_y}  U (\rho \otimes \ket{\phi}\bra{\phi}) U^\dagger \hat{P_y}^\dagger dy
\end{eqnarray}

\noindent or

\begin{eqnarray}
\mathcal{E}_{\hat{A}}^W(\rho) = \int p(y) \rho_y dy
\end{eqnarray}

\noindent where

\[ \rho_y = \frac{\hat{P}_y  U (\rho \otimes \ket{\phi}\bra{\phi}) U^\dagger \hat{P}_y^\dagger} {tr(\hat{P}_y  U (\rho \otimes \ket{\phi}\bra{\phi}) U^\dagger \hat{P}_y^\dagger)} \]

\noindent and

\[ p(y) = tr(\hat{P}_y  U (\rho \otimes \ket{\phi}\bra{\phi}) U^\dagger \hat{P}_y^\dagger) .\]

Bob is now using a POVM measurement. By the HSW theorem, the rate of transmission in the channel between Alice and Bob after the weak eavesdropping is:

\begin{eqnarray}
\chi(\mathcal{E}_{\hat{A}}^W ) = \max_{p_i,\rho_i} \{ S(\mathcal{E}_{\hat{A}}^W(\rho)) - \sum p_i S( \mathcal{E}_{\hat{A}}^W(\rho_i))\}.
\end{eqnarray}

We can compare this capacity with the capacity of the same channel without eavesdropping. The reduction in the capacity will be the result of the information `leaking' through the weak channel between Alice and Eve. In fact we can estimate this reduction using the capacity of the channel between Alice and Eve.

In the above quantum weak Gaussian channels we impose restrictions on the measurement at the end point of the channel; a weak measurement using an Hermitian operator $\hat{A}$. In general if Alice and Bob could use any Hermitian operator the capacity would be:

\[ \max_{\hat{A}} \max_{X(\hat{A})} H(X:X+Z)\]

\noindent where the first maximum is taken over all Hermitian operator with distinct eigenvalues.


There is reason to believe that one could also extend other theorems on classical Gaussian channels to the quantum case, such as the `water filling' theorem \cite{Cover}: given a set of quantum weak Gaussian channels between Alice and Bob, each with its own capacity, what will be the order of usage of the channels to maximize the rate of information passage between them.\\

We hope the above results will open the way to further analysis of weak measurements and better understanding of their pros and cons.

\section{Acknowledgements}

We wish to thank two anonymous referees for many helpful remarks. E.C. was partially supported by Israel Science Foundation Grant No. 1311/14.



\end{titlepage}

\begin{thebibliography}{99}


\bibitem{Shannon} C.E. Shannon, W. Weaver, The mathematical theory of communication, Univ. of Illinois Press (1998).

\bibitem{Cover} T. Cover, C. Thomas, Elements of information theory, John Wiley and Sons, New-York (1991).

\bibitem{HolevoG2}
A.S. Holevo, M. Sohma, O. Hirota. Capacity of quantum Gaussian channels, Phys. Rev. A 59.3 (1999): 1820-1828.‏

\bibitem{HolevoG1}
A.S. Holevo, Coding theorems for quantum channels, arXiv preprint quant-ph/9809023 (1998).


\bibitem{G3}
A.S Holevo, and R.F. Werner, Evaluating capacities of bosonic Gaussian channels, Phys. Rev. A 63.3 (2001): 032312.


\bibitem{G4}
J. Eisert, M.M. Wolf, Gaussian quantum channels, arXiv preprint quant-ph/0505151 (2005).‏

\bibitem{G5}
A.S Holevo, One-mode quantum Gaussian channels: Structure and quantum capacity, Problems of Information Transmission 43.1 (2007): 1-11.‏‏

\bibitem{AAV}
Y. Aharonov, D. Albert, L. Vaidman, How the result
of a measurement of a component of the spin of a spin-1/2 particle
can turn out to be 100, Phys. Rev. Lett. 60.14 (1988): 1351-1354.


\bibitem{Aharonov}
Y. Aharonov, D. Rohrich, Quantum Paradoxes: Quantum
Theory for the Perplexed. Wiley-VCH (2005).

\bibitem{Tamir}
B. Tamir, E. Cohen, Introduction to weak measurements and weak values, Quanta 2.1 (2013): 7-17.

\bibitem{ACE}
Y. Aharonov, E. Cohen, A.C.  Elitzur, Foundations and applications of weak quantum measurements, Phys. Rev. A 89.5 (2014): 052105.

\bibitem{Nielsen}
M.A. Nielsen, I.L. Chuang, Quantum computation and quantum information, Cambridge university press (2000).

\bibitem{SW97}
B. Schumacher, M.D. Westmoreland, Sending classical information via noisy quantum channels, Phys. Rev.  A 56.1 (1997): 131.



\bibitem{Neumann} J. von Neumann, Mathematical Foundations of Quantum
Mechanics. Investigations In Physics, Beyer RT (translator), Princeton:
Princeton University Press (1955).

\bibitem{Peres} A. Peres, Quantum theory: Concepts and methods, Kluwer (2002).




\bibitem{Vaidman} L. Vaidman, Weak-measurement elements of reality, Found. Phys. 26.7 (1996): 895-906.


































































































































\end{thebibliography}
\end{document}